# Anisotropic Energy Gaps of Iron-based Superconductivity from Intra-band Quasiparticle Interference in LiFeAs

M. P. Allan[§ 1,2,3], A. W. Rost[§ 2,3], A. P. Mackenzie[3], Yang Xie[2], J. C. Davis[1,2,3,4]✝, K. Kihou[5,6], C. H. Lee[5,6], A. Iyo[5,6], H. Eisaki[5,6], and T.-M. Chuang[1,2,7]✝

[1] Condensed Matter Physics and Materials Science (CMPMS) Department, Brookhaven National Laboratory, Upton, NY 11973, USA

[2] Laboratory of Solid State Physics, Department of Physics, Cornell University, Ithaca, NY 14853, USA.

[3] Scottish Universities Physics Alliance (SUPA), School of Physics and Astronomy, University of St Andrews, St Andrews, Fife KY16 9SS, UK.

[4] Kavli Institute at Cornell for Nanoscale Science, Cornell University, Ithaca, NY 14853, USA.

[5] Institute of Advanced Industrial Science and Technology, Tsukuba, Ibaraki 305-8568, Japan.

[6] Japan Science and Technology Agency (JST), Transformative Research-Project on Iron Pnictides (TRIP), Tokyo 102-0075, Japan.

[7] Institute of Physics, Academia Sinica, Nankang, Taipei 11529, Taiwan.

§ These authors contributed equally to this work.
✝To whom correspondence should be addressed. E-mail: jcdavis@ccmr.cornell.edu (J.C.D.); chuangtm@phys.sinica. edu.tw (T.M.C.)

**1       If strong electron-electron interactions between neighboring Fe atoms mediate the Cooper pairing in iron-pnictide superconductors, then specific and distinct anisotropic superconducting energy gaps $\Delta_i(\vec{k})$ should appear on the different electronic bands *i*. Here we introduce intra-band Bogoliubov quasiparticle scattering interference (QPI) techniques for determination of $\Delta_i(\vec{k})$ in such materials, focusing on LiFeAs. We identify the three hole-like bands assigned previously as $\gamma$, $\alpha_2$ and $\alpha_1$, and we determine the anisotropy, magnitude and relative orientations of their $\Delta_i(\vec{k})$. These measurements will advance quantitative theoretical analysis of the mechanism of Cooper pairing in iron-based superconductivity.**

2      In typical FeAs-based materials, every second As atom lies above/below the FeAs layer (Fig. 1A) so that the crystallographic unit cell, instead of being a square with an Fe atom at each corner (Fig. 1A, dashed box), is rotated by 45° and has an As at each corner (Fig. 1A, solid box). The corresponding momentum space ($\vec{k}$-space) Brillouin zone (BZ) then contains five electronic bands; the hole-like $\alpha_1$, $\alpha_2$ and $\gamma$ bands surround the $\Gamma$ point, and the electron-like $\beta_1$ and $\beta_2$ bands surrounding the $\tilde{M}$ point (Fig. 1B). The superconductivity derives (*1*) from a commensurate-antiferromagnetic and orthorhombic 'parent' state (see supplementary materials *(2)*). The highest superconducting critical temperatures ($T_c$) occur when the magnetic/structural transitions are suppressed towards zero temperature. Theories describing the FeAs superconductivity can be quite complex (*1,3-9*) but they typically contain two essential ingredients: (i) the predominant superconducting order parameter (OP) symmetry is *s±*, i.e. has *s*-wave symmetry but changes sign between different bands, and, (ii) the superconducting energy gap functions $\Delta_i(\vec{k})$ on different bands *i* are anisotropic in $\vec{k}$-space, with each exhibiting distinct 90°-rotational ($C_4$) symmetry and a specific relationship of gap minima/maxima relative to the BZ axes. Figure 1C shows a schematic of such a situation for just two electronic bands, with contours of constant energy (CCE) for their Bogoliubov quasiparticles shown in Fig. 1D.

3      Although there exists evidence for *s±* OP symmetry (*10,11*), the structure of any anisotropic gaps $\Delta_i(\vec{k})$ on different bands and their relative $\vec{k}$-space orientation is an open question for virtually all iron-based superconductors (*1*). However, it is the structure of these $\Delta_i(\vec{k})$ that is crucial for understanding the pairing interactions. Thermodynamic and transport studies (which cannot reveal $\Delta_i(\vec{k})$) provide good evidence for electronic anisotropy (*1,12,13,14*). By contrast, almost all $\vec{k}$-space resolved photoemission spectroscopy (ARPES) studies of these materials, including LiFeAs (*15*), have reported the $\Delta_i(\vec{k})$ to be isotropic in the $k_x/k_y$ plane (*1*). Subsequent to our submission, however, nodeless anisotropic gaps with values spanning the range 2mV→4meV on the γ band and 5meV→6meV on the $\alpha_2$ band (surrounding Γ) were reported in LiFeAs (*16,17*). For the electron-like pocket at $\widetilde{M}$, an anisotropic 3meV→4.5meV gap is also described, but the reported $\vec{k}$-space positions of gap maxima/minima appear mutually inconsistent (*16,17*). High-resolution determination of $\Delta_i(\vec{k})$ should help to more accurately quantify these fundamental characteristics of the superconductivity.

4    Bogoliubov quasiparticle scattering interference (QPI) imaging is a suitable technique for high-resolution determination of $\Delta_i(\vec{k})$ (*18-24*). The scattering interference patterns can be visualized in real space ($\vec{r}$-space) using spectroscopic imaging scanning tunneling microscopy (SI-STM) in which the tip-sample differential tunneling conductance $dI/dV(\vec{r},E) \equiv g(\vec{r},E)$ is measured as a function of location $\vec{r}$ and electron energy $E$. However, using QPI to determine the $\Delta_i(\vec{k})$ of iron-pnictide superconductors may be problematic because: (i) the large field-of-view $g(\vec{r},E)$ imaging (and equivalent high $\vec{q}$-space resolution in $g(\vec{q},E)$) necessary to obtain $\Delta_i(\vec{k})$ is technically difficult; (ii) complex overlapping QPI patterns are expected from multiple bands, (iii) most iron-pnictide compounds exhibit poor cleave-surface morphology. This latter point can be mitigated by using a material with a charge-neutral cleave plane (e.g. *11*). An iron-pnictide that satisfies this requirement is LiFeAs (*25-28*) which has a glide plane between two Li layers (*(2)*, section I).

5    Bogoliubov QPI can be influenced by a variety of effects in the iron-pnictides (*21-24*). To explore the expected QPI signatures of the $\Delta_i(\vec{k})$, we consider the model two-gap structure in Fig. 1C and D. Within one band, each energy $\Delta_1^{min} \leq E \leq \Delta_1^{max}$ at which $\Delta_1(\vec{k}) \doteq E$ picks out eight specific $\vec{k}$-space locations $\vec{k}_j(E)$. In a generalization of the 'octet' model of QPI in copper-based superconductors (*18-20*), scattering between these $\vec{k}_j(E)$ should produce interference patterns with the seven characteristic QPI wavevectors $\vec{q}_1....\vec{q}_7$ shown in Fig. 1D. An equivalent set of QPI wavevectors, but now with different lengths/orientations, would be generated by a different $\Delta_2(\vec{k})$ on the second band (Fig. 1C,D, yellow). Thus, conventional 'octet' analysis could be challenging for a multi-band anisotropic superconductor, as many intra-band QPI wavevectors coincide near the center of $\vec{q}$-space.

6       To achieve a more robust QPI prediction we next consider the joint-density-of-states (JDOS) approximation for the whole Bogoliubov quasiparticle spectrum (Fig. 1D). In this case, for $\Delta_1^{\min} \leq E \leq \Delta_1^{\max}$ within a given band (e.g. red band in Fig. 1C,D), the tips of the 'bananas' still strongly influence QPI because of the enhanced JDOS for scattering between these locations and effects inherent to the structure of the coherence factors. Figure 1E shows $g(\vec{q},E=(\Delta_{\max}+\Delta_{\min})/2)$ simulated by calculating the JDOS for a single hole-like band (red band of Fig. 1C, see *(2)*, section II). Although the JDOS approximation gives an intuitive picture of how $g(\vec{q},E)$ relates to regions of high density of states in $\vec{k}$-space, the T-matrix formalism is needed for a rigorous description (*18-24*). In Fig. 1F we show a T-matrix simulation of $g(\vec{q},E=(\Delta_{\max}+\Delta_{\min})/2)$ for equivalent parameters as those in Fig. 1E (*(2)*, section II). By comparison of Fig. 1E and 1F we see that the two types of simulations are virtually indistinguishable, and that the expected octet wavevectors (overlaid by black arrows) are in excellent agreement (as is true for all $\Delta_1^{\min} \leq E \leq \Delta_1^{\max}$). The key QPI signatures of an anisotropic but nodeless $\Delta_i(\vec{k})$ are therefore expected to be arcs of strong scattering centered along the direction of the gap minima with regions of minimal scattering intensity located towards the gap maxima (green arrows Fig. 1E,F).

7       The complexity of the simulated $g(\vec{q},E)$ (Fig. 1E,F) also reveals the considerable practical challenge if measured $g(\vec{q},E)$ are to be inverted to yield the underlying $\Delta_i(\vec{k})$. Therefore we developed a restricted analysis scheme that still allows the pertinent $\Delta_i(\vec{k})$ information to be easily extracted. Instead of the usual constant energy $g(\vec{q},E)$ images, this approach is based on measuring the maximum scattering intensity in a $|\vec{q}|$-$E$ plane along a specific high-symmetry direction, for example $\vec{q} \parallel \Gamma\widetilde{M}$ for the red band in Fig. 1D. Fig. 1G shows a simulation of such a $|\vec{q}|$-$E$ intensity plot ($\vec{q} \parallel \Gamma\widetilde{M}$) revealing two curved trajectories of maximal scattering intensity. They are extracted and plotted as red curves in Fig. 1H (*(2)*, section III). Such a plot of intensity maxima in a $g(|\vec{q}|,E)$ plane along a high symmetry direction actually contains all the information on $\Delta(\vec{k})$ for that $C_4$-symmetric band. Another band with a different $\vec{q}$-space radius and whose gap maxima are rotated by 45° to the first (e.g. yellow in Fig. 1D) can be analyzed similarly; a plot of maxima in $g(|\vec{q}|,E)$ for $\vec{q} \parallel \Gamma\widetilde{X}$ (yellow arrow Fig. 1D) would then yield distinct interference maxima from the $\Delta(\vec{k})$ of that band (yellow curves Fig. 1H). The magnitude and relative orientation of $\Delta_i(\vec{k})$ on multiple $C_4$-symmetric bands of different $|\vec{q}|$ radius can be determined simultaneously by using this multiband QPI approach. Importantly, the $\Delta_i(\vec{k})$ are then determined experimentally, not from comparison to simulation, but directly from a combination of: (i) the measured normal-state band dispersions, (ii) the BZ symmetry, and (iii) the measured geometrical characteristics of the scattering intensity $g(|\vec{q}|,E)$ in a specific $|\vec{q}|$-$E$ plane (*(2)*, section III).

8   We implement this approach using LiFeAs crystals with $T_c \approx 15$K. Their cleaved surfaces are atomically flat (Fig. 2A) and exhibit the $a_0$=0.38nm periodicity of either the As or Li layer (*(2)*, section I). As the glide plane is between Li layers we are probably observing a Li-termination layer. We image $g(\vec{r},E)$ with atomic resolution and with a thermal energy-resolution $\delta E \leq 350$ μeV at $T$=1.2K. Fig. 2A inset shows the spatially averaged differential conductance $g(E)$ far from in-gap impurity states (*(2)*, section V). The density of electronic states $N(E) \propto g(E)$ (Fig. 2A, inset) is in agreement with the $N(E)$ first reported for LiFeAs (*29, 30*). It indicates: (i) a fully gapped superconductor because $g(E)$ ~0 for |E|<1meV, (ii) a maximum energy gap of ~6meV and, (iii) a complex internal structure to $N(E)$ consistent with strong $\vec{k}$-space gap anisotropy.

9    Next we image $g(\vec{r},E)$ in ~90 nm square fields of view (FOV), at temperatures between 1.2K and 16K, and in the energy range associated with Cooper pairing (*(2)*, section VI). The large FOV is required to achieve sufficient $\vec{q}$-space resolution. Figures 2B, C show a typical $g(\vec{r},E)$ and its Fourier transform $g(\vec{q},E)$ with $E$=7.7 meV, and the inset shows a typical $\vec{r}$-space example of the interference patterns. Figures 2C,D show representative Fourier transforms $g(\vec{q},E)$ from energies above the maximum superconducting gap magnitude ($E$=+7.7 meV, -6.6 meV). The scattering interference signatures for three distinct bands can be detected as closed contours in $\vec{q}$-space; we refer to them as $h_3, h_2, h_1$ throughout. The measured $|\vec{q}(E)|$ of these bands in Fig. 2E shows them all to be hole-like. A quantitative comparison of QPI to both ARPES (*15-17*) and quantum oscillation measurements (*(31), (2)*, section VII) identifies $h_1$, $h_2$ and $h_3$ with the three hole-like bands assigned $\alpha_1$, $\alpha_2$ and $\gamma$. The QPI signatures of electron-like bands are weak and complex, perhaps because they share the same $\vec{q}$-space regions as the signature of $h_2$, or because of the stronger $k_z$ dispersion of these bands, or because of weak overlap between high-$|\vec{k}|$ states and the localized tip-electron wavefunction; we do not consider them further here.

10    Figures 3A,B show the $g(\vec{q},E)$ measured within the energy gaps in Fig. 2E. Scattering interference is virtually nonexistent in the $\vec{q}$-space direction parallel to $\Gamma\widetilde{X}$ for the $h_3$ band (Fig. 3A, green arrows), and is similarly faint in a direction parallel to $\Gamma\widetilde{M}$ for the $h_2$ band (Fig. 3B, green arrows). The disappearance in the superconducting phase of QPI in these two directions is a result of an anisotropic gap opening at relevant $\vec{k}$-space locations on these bands (see insets Fig. 3A,B), as predicted from both the T-matrix and JDOS simulations in Fig. 1E,F. Fig. 3C shows $g(\vec{\theta},E)$, the strongly anisotropic scattering intensity versus angle in $\vec{q}$-space $g(\vec{\theta},E)$ on the $h_3$ Fermi surface. The $g(\vec{q},E)$ data in Fig. 3A,B, together with the $g(\vec{\theta},E)$ data in Fig. 3C reveal unambiguous QPI signatures for anisotropic gaps of different orientations on different bands in LiFeAs.

11      To quantify the $\Delta_i(\vec{k})$, we examine the scattering intensity in $g(\vec{q},E)$ within a $|\vec{q}|$-$E$ plane defined first by $\vec{q}$ parallel to $\Gamma\widetilde{M}$, and then by $\vec{q}$ parallel to $\Gamma\widetilde{X}$. The locus of maximum intensity in the $g(|\vec{q}|,E)$ plane with $|\vec{q}|$ parallel to the $\Gamma\widetilde{M}$ direction is extracted (*(2)*, section VIII) and plotted in Fig. 3D. The evolution of the normal-state $h_3$ band (red in Fig. 2E) is indicated here by red arrows. The superconducting energy gap minimum of $h_3$ is seen, along with the curved trajectories of scattering intensity maxima expected from an anisotropic $\Delta(\vec{k})$ (Fig. 1G,H and *(2)*, section VIII). Similarly, by measuring $g(|\vec{q}|,E)$ in the plane with $|\vec{q}|$ parallel to $\Gamma\widetilde{X}$, the locus of maximum intensity of the $h_2$ band is extracted (*(2)*, section VIII) and plotted in Fig. 3E. The energy gap minimum is again obvious along with the curved scattering intensity maxima expected from a different anisotropic $\Delta(\vec{k})$ (see Fig. 1H); the data for the upper branch cannot be obtained because of interference from other signals (possibly the electron-like band). The QPI signature of the third band $h_1$ becomes unidentifiable within -6±1.5 meV below $E_F$ in the superconducting phase, consistent with the opening of a gap of this magnitude (Figs. 2E and 3E) but we cannot yet resolve any gap modulations.

12      The magnitude, anisotropy and relative position of $\Delta_i(\vec{k})$ on bands $h_3$, $h_2$, $h_1$ are then determined from Fig. 3D,E using the previously described procedure (*(2)*, section III). The resulting anisotropic superconducting gaps on bands $h_3$, $h_2$, $h_1$ of LiFeAs are displayed in Fig. 4A and Fig. 4B. While our $g(|\vec{q}|,E)$ agree well with pioneering QPI studies of LiFeAs where common data exist, no studies of $\Delta_i(\vec{k})$ were reported therein (*30*). Moreover, while field-dependent Bogoliubov QPI can reveal OP symmetry (*11*), these techniques were not applied to LiFeAs and no OP symmetry conclusions drawn herein. The anisotropic $\Delta_i$ reported recently in ARPES studies of LiFeAs (*16,17*) appear in agreement with our observations for the $h_3$ (Fig. 1, γ) and $h_2$ (Fig. 1 α$_2$) bands. Lastly our measurements are quite consistent with deductions on LiFeAs band structure from quantum oscillation studies (*31*). Overall, the growing confidence and concord in the structure of $\Delta_i(\vec{k})$ for LiFeAs will advance the quantitative theoretical study of the mechanism of its Cooper pairing. Moreover, the multi-band anisotropic-gap QPI techniques introduced here will allow equivalent $\Delta_i(\vec{k})$ observations in other iron-pnictide superconductors.

**Figure 1**

(A) Top view of crystal structure in the FeAs plane. Dashed lines represent the one-Fe unit cell that would exist if all As were coplanar, while the actual unit cell of dimension $a_0 \approx 0.38$ nm is shown using solid lines.

(B) Schematic Fermi surface of an iron-based superconductor like LiFeAs in the tetragonal non-magnetic Brillouin zone (solid line). The 'one Fe zone' is shown as a dashed line. The blue, yellow and red curves show the hole-like pockets surrounding the $\Gamma$ point, while the green curves show electron pockets surrounding the $\widetilde{M}$ point. The grey crosses mark the $(\pm 1/2, \pm 1/2)\pi/a_0$ points.

(C) Model system exhibiting two distinct anisotropic energy gaps $\Delta_i(\vec{k})$ on two hole-like bands.

(D) Contours-of-constant-energy (CCE) of the Bogoliubov quasiparticle excitation spectrum for the two bands in (C), each in the same color as its $\Delta_i(\vec{k})$. Contours enclose diminishing areas surrounding the gap minimum in each case. Black arrows indicate the 'octet' scattering vectors $\vec{q}_1 \ldots \vec{q}_7$ between the CCE 'banana tips' (red dots). The red and yellow vectors indicate the high-symmetry directions along which $g(|\vec{q}|,E)$ should be measured to determine $\Delta_i(\vec{k})$ for the respective bands separately – see (H).

(E,F) Theoretically simulated $g(\vec{q},E=(\Delta^{max}+\Delta^{min})/2))$ for a single hole-like band with band- and gap-anisotropy parameters shown in D, using the JDOS (E), and T-matrix (F) approaches. In $\vec{q}$-space, the grey crosses occur at the $(\pm 1/2, \pm 1/2)2\pi/a_0$ points. The arrows show the scattering vectors $\vec{q}_1 \ldots \vec{q}_7$ from the octet model in (D). The red dots indicate the vectors which lie along $\Gamma\widetilde{M}$, the direction studied in (G) and (H)

(G) JDOS simulation of dependence of scattering intensity in $g(|\vec{q}|,E)$ with $\vec{q}$ parallel to $\Gamma\widetilde{M}$ for a single hole-like band with gap anisotropy parameters shown in (C,D). Note the curved shapes of scattering intensity maxima which we then extract in (H).

(H) Red curves: Expected trajectory of maxima in scattering intensity $g(|\vec{q}|,E)$ with $\vec{q}$ parallel to $\Gamma\tilde{M}$ for $h_3$. Yellow curves: Expected maxima in scattering intensity $g(|\vec{q}|,E)$ with $\vec{q}$ parallel to $\Gamma\tilde{X}$ for $h_2$.

**Figure 2**

(A) A ~35 nm square topographic image of LiFeAs surface, taken at $V_{Bias}$=-20mV, $R_{Junction}$=2 G$\Omega$, $T$=1.2K. Inset: average spectrum over an area without impurities (see (2), section V). The maximum gap superconducting coherence peaks and the zero conductance near $E_F$ are clear.

(B) Typical differential conductance image $g(\vec{r},E$=7.7meV) at $T$=1.2K. Inset: typical QPI oscillations in $g(\vec{r},E)$ surrounding an impurity atom.

(C) The $g(\vec{q},E$=7.7meV) which is the power-spectral-density Fourier transform of $g(\vec{r},E$=7.7meV) from (B). The high intensity closed contour is a consequence of scattering interference within a large hole-like band $h_3$.

(D) The $g(\vec{q},E$=-6.6meV) showing the high intensity closed contours resulting from scattering interference from a smaller hole-like band $h_2$, and $h_1$ (inset). These data are measured at $T$=1.2K.

(E) The measured energy dependence of the $\vec{q}(E)=2\vec{k}(E)$ for all three bands $h_1$ (blue line), $h_2$ (yellow) and $h_3$ (red) along the marked directions. Black dots are measured in the superconducting state at $T$=1.2K while gray dots are measured in the normal state at $T$=16K.

**Figure 3**

(A) $g(\vec{q},E$=2meV) measured at $T$=1.2K. The scattering on the $h_3$ band has become highly anisotropic within its energy gap range with vanishing intensity in the $\Gamma\tilde{M}$ direction (green arrows). The inset shows schematically in $\vec{k}$-space how this is indicative of a lower energy gap along $\Gamma\tilde{M}$ and a higher energy gap along $\Gamma\tilde{X}$.

(B) $g(\vec{q},E$=-5meV) measured at $T$=1.2K. The scattering on the $h_2$ band has become highly anisotropic within its energy gap range (green arrows), indicative of a lower energy gap along $\Gamma\tilde{X}$ and a higher energy gap along $\Gamma\tilde{M}$.

(C) A projection of the QPI intensity $g(\vec{q},E)$ (a three-dimensional dataset) onto the measured normal-state band dispersion $\varepsilon(\vec{k}) = \varepsilon(\vec{q}/2)$. In the direction of minimal gap ($\Gamma\tilde{M}$) the intensity is gapped up to around 2meV $\approx \Delta_{h3}^{min}$, whereas in the direction of maximal gap ($\Gamma\tilde{X}$), it is gapped up to around 3meV $\approx \Delta_{h3}^{max}$.

(D) Extracted maximum scattering intensity trajectory from $g(|\vec{q}|,E)$ for $\vec{q} \parallel \Gamma\tilde{M}$ ((2), section VIII) containing the information on $\Delta(\vec{k})$ for $h_3$.

(E) Extracted maximum scattering intensity trajectory from $g(|\vec{q}|,E)$ for $\vec{q} \parallel \Gamma\tilde{X}$ ((2), section VIII) containing the information on $\Delta(\vec{k})$ for $h_2$. Blue: Gap opening on the $h_1$ band.

**Figure 4**

(A) Anisotropic energy gap structure $\Delta_i$ measured using QPI at $T$=1.2K on the three hole-like bands $h_3$, $h_2$ and $h_1$ (Fig. 2E). These bands have been labeled $\gamma$, $\alpha_1$ and $\alpha_2$ before. Here, the 0.35mV error bars stem from the thermal resolution of SI-STM at 1.2K.

(B) A three-dimensional rendering of the measured (solid dots) anisotropic energy gap structure $\Delta_i$ on the three hole-like bands at $T$=1.2K.

**Acknowledgements:** We are particularly grateful to D.-H. Lee for advice and discussions and we acknowledge and thank F. Baumberger, P.C. Canfield, A. Carrington, A.V. Chubukov, A. I. Coldea, M.H. Fischer, T. Hanaguri, P.J. Hirschfeld, B. Keimer, E.-A. Kim, M.J. Lawler, C. Putzke, J. Schmalian, H. Takagi, Z. Tesanovic, R. Thomale, S. Uchida and F. Wang for helpful discussions and communications. Studies were supported by the Center for Emergent Superconductivity, an Energy Frontier Research Center, funded by the U.S. Department of Energy under DE-2009-BNL-PM015; by the UK Engineering and Physical Sciences Research Council (EPSRC); by a Grant-in-Aid for Scientific Research C (No. 22540380) from the Japan Society for the Promotion of Science. Y.X. acknowledges support by the Cornell Center for Materials Research (CCMR) under NSF/DMR-0520404. T.-M.C. acknowledges support by Academia Sinica Research Program on Nanoscience & Nanotechnology, and APM the receipt of a Royal Society-Wolfson Research Merit Award. The data described in the paper are archived by the Davis Research Group at Cornell University.

**Supporting Online Material**

Supplementary Text

Figs. S1 to S10

References *(32-36)*

*Figure 1*

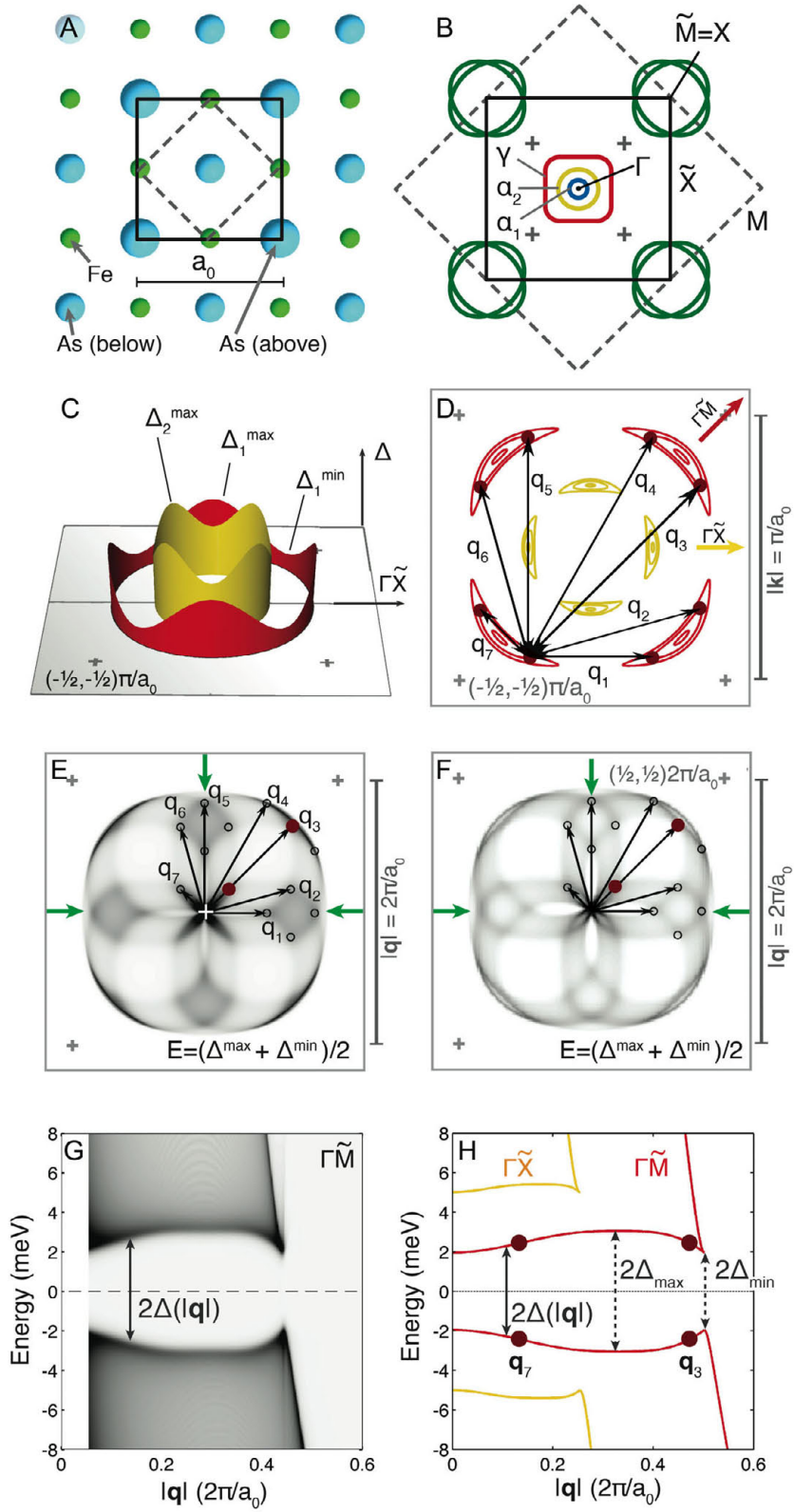

*Figure 2*

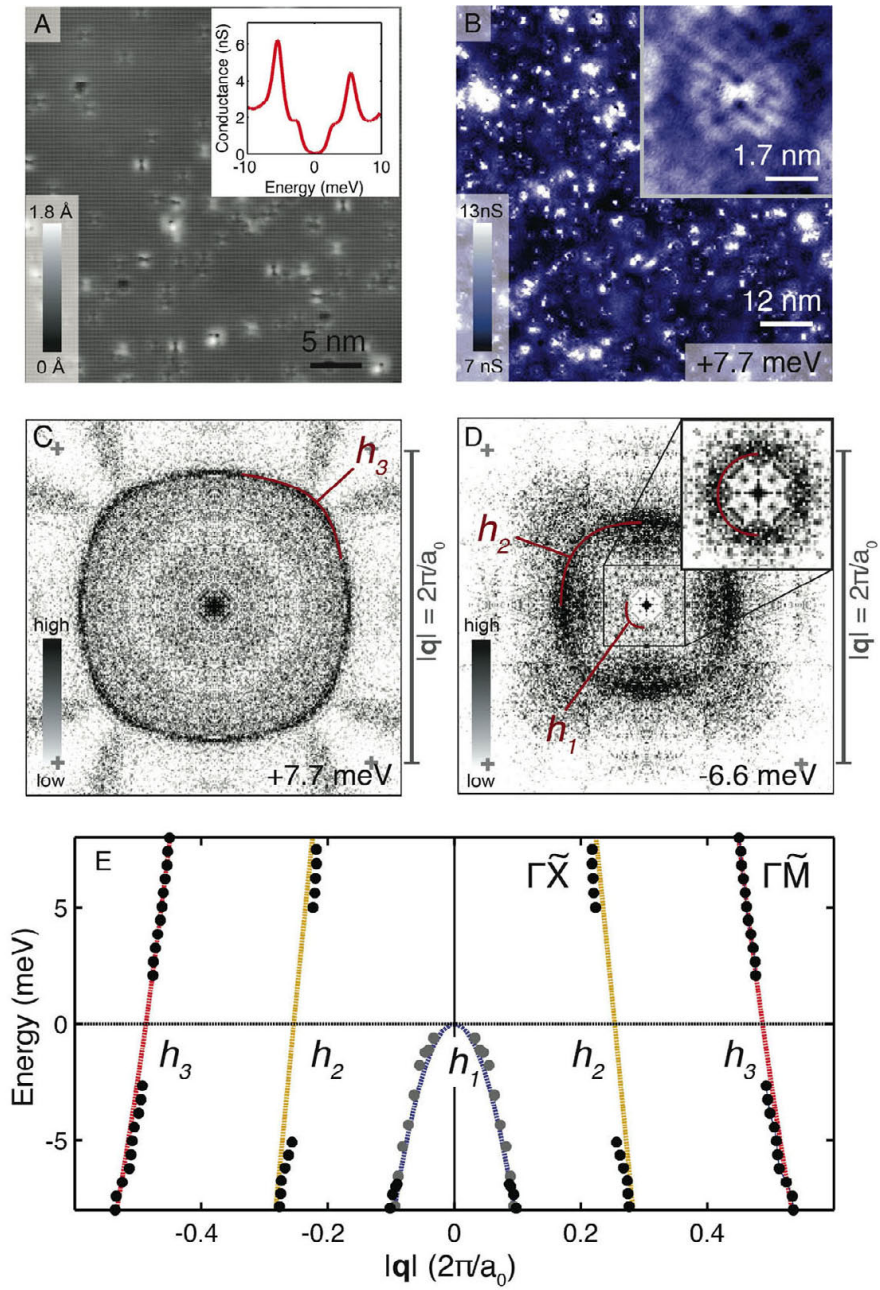

*Figure 3*

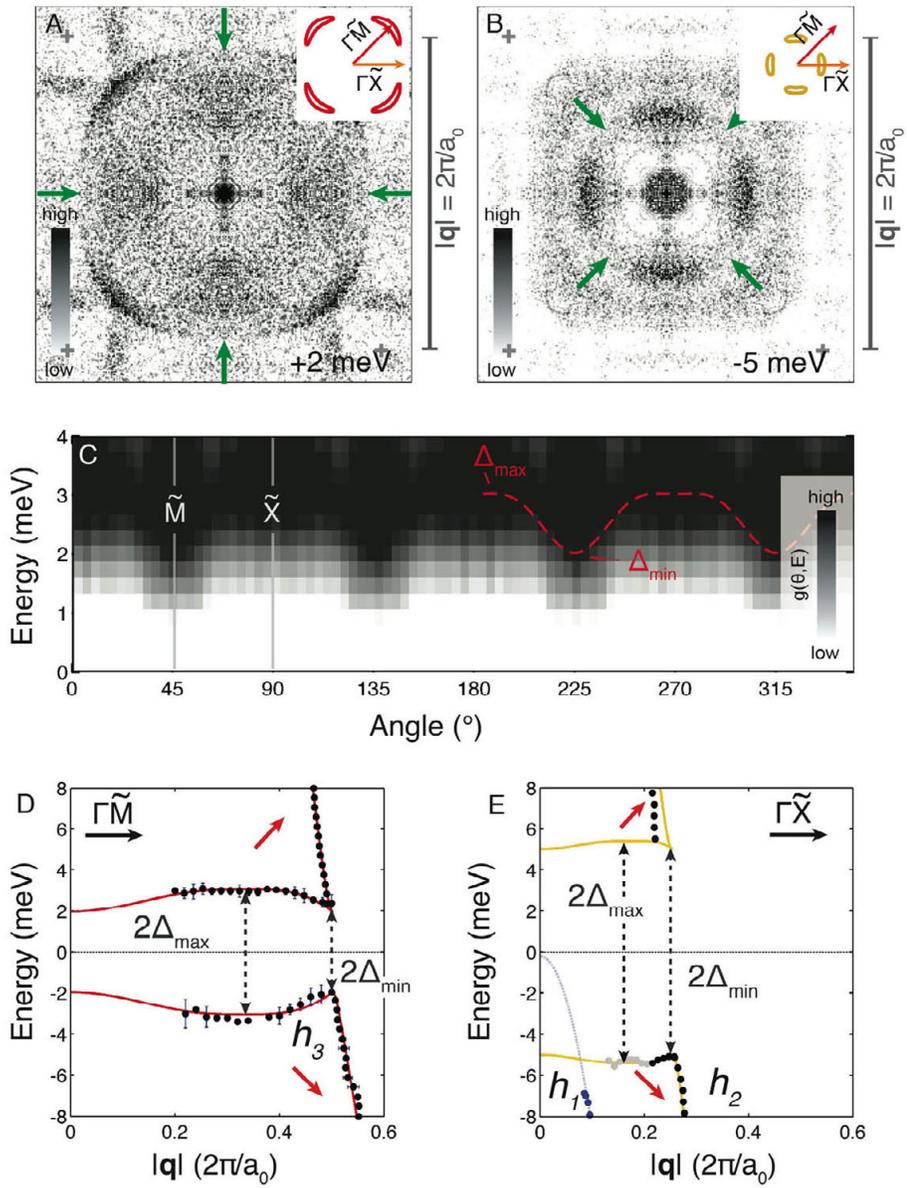

*Figure 4*

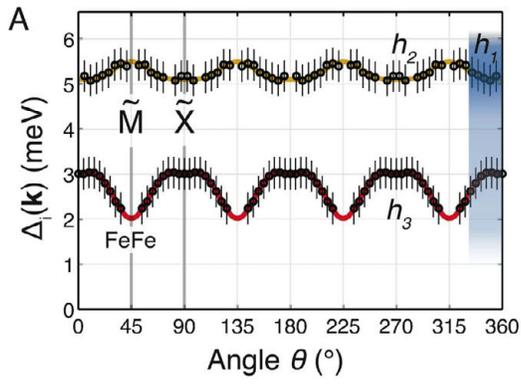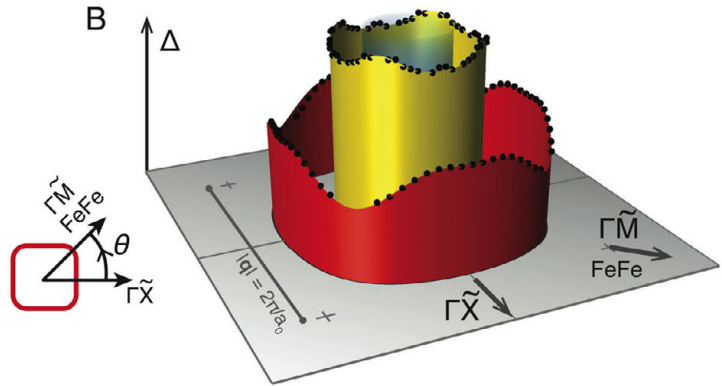

Supporting Online Material for

*Anisotropic Energy-Gaps of Iron-based Superconductivity from Intra-band Quasiparticle Interference in LiFeAs*

M. P. Allan*, A. W. Rost*, A. P. Mackenzie, Yang Xie, J. C. Davis,
K. Kihou, C.-H. Lee, A. Iyo, H. Eisaki, and T.-M. Chuang.

* These authors contributed equally to this work.

*This PDF file includes:* SOM Text Figs. S1 to S10 References

# Table of Contents

# I. Materials and Methods A

We use LiFeAs crystals prepared by the LiAs flux method yielding $T_c \approx 15K$. The highly reactive samples are prepared in a dry-nitrogen atmosphere and cleaved at low temperatures in cryogenic ultra-high-vacuum. All of the QPI measurements reported here were performed between $T=1.2K$ and $T=16K$.

## I.1. *Crystal Structure of the sample and Cleave Plane*

Most iron-pnictides do not posses a charge neutral cleave plane. This is likely the reason for the surface reconstructions that form on many surfaces. In LiFeAs, however, the FeAs layers are separated by two Li layers, with a glide plane between. Thus there exists a charge-neutral plane where cleaving is likely to occur. Moreover, it has been shown that no surface reconstruction occurs on such a cleaved surface (34).

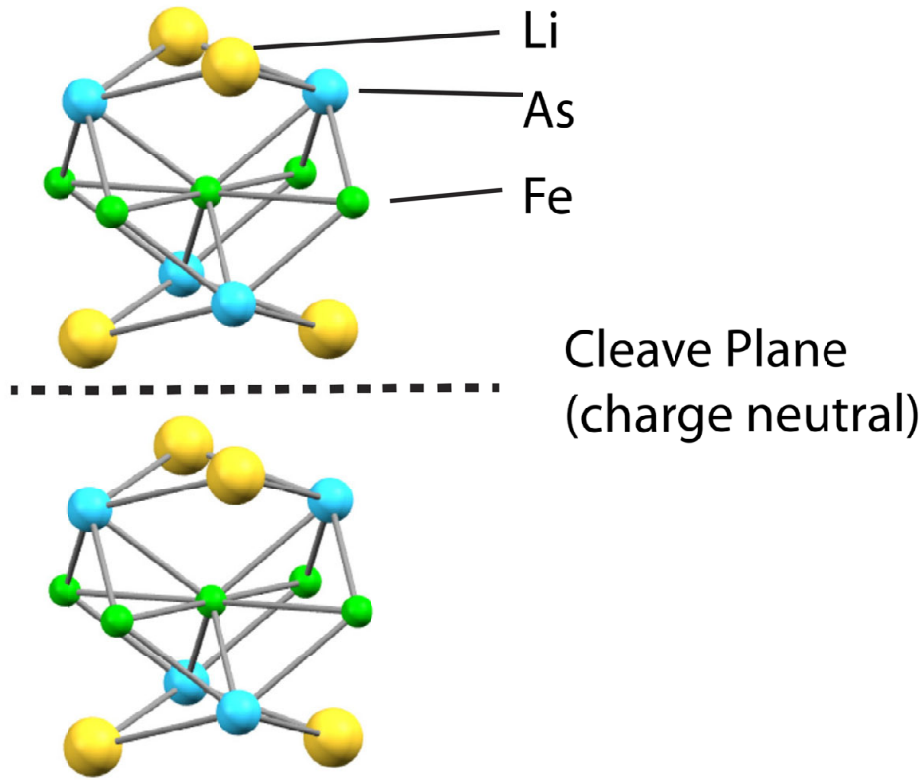

**Figure S1** | Crystal structure and likely cleave plane.

*I.2.    Magnetic properties of the sample*

Many Fe-based superconductors exhibit long-range magnetic order in their parent states. The spins are aligned along one Fe-Fe direction, and anti-aligned in the other direction (Fig. S2 shows a FeAs layer with the typical spin orientations). This magnetic order and orthorhombicity can be suppressed by chemical substitution or pressure, but even then, gapped spin fluctuations with the same wave-vectors exist. Indeed it is these fluctuations that are believed by many to be relevant for the mechanism of superconductivity in these compounds (*1,3-9*).

Stoichiometric LiFeAs seems not to exhibit long-range magnetic order. However, recent inelastic neutron scattering experiments show strong spin fluctuations (32,33) down to very low energies and with wave vectors that are consistent with the spin-density wave structure in other iron-based superconductors.

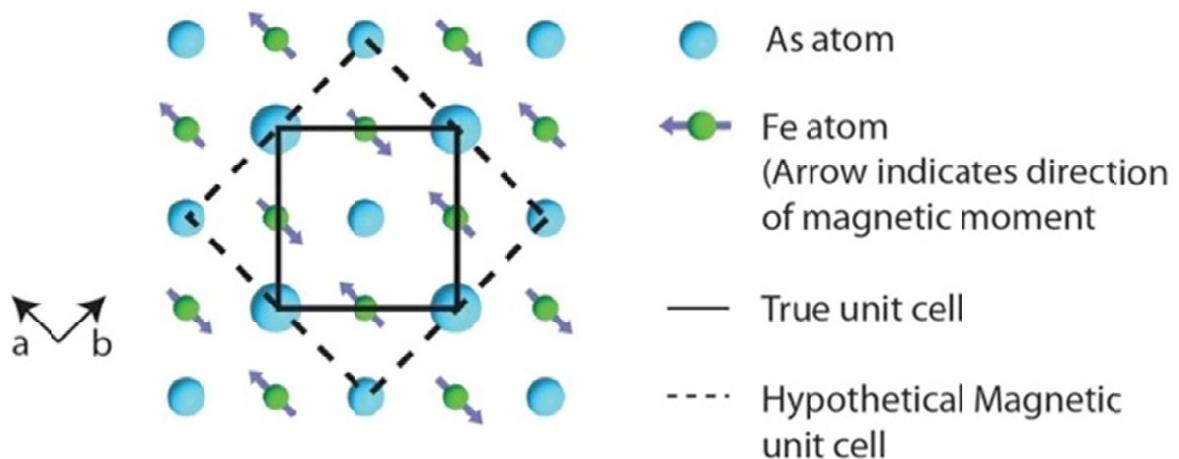

**Fig. S2 | Real space structure of antiferromagnetic order.** Top view of crystal structure in the FeAs plane; blue and green circles represent Fe and As atoms respectively with As atoms above/below the plane being represented by larger/smaller spheres. The size of the true LiFeAs unit cell is shown as a solid line with the dashed line representing the size of the magnetic unit cell, typical of the parent state in other FeAs compounds.

## II. Simulation of $g(\vec{q}, E)$ for a Single Hole-Like Band using the $\hat{T}$-Matrix and JDOS Approaches, and Comparison to Octet-Model

What is observed in $g(\vec{q}, E)$ maps derived from Fourier transform studies of Spectroscopic Imaging-STM (SI-STM) conductance images $g(\vec{r}, E)$, are effects due to the static screening response of the electronic liquid to a local perturbation.

Bogoliubov QPI in iron-pnictides can, in theory, be influenced by several effects (*9,18-24*). First, the Bogoliubov coherence factors of initial and final states make the interference patterns sensitive to the relative particle-hole mixing as well as the potential/magnetic component of the scattering potential. Second, scattering between states in bands which have different orbital decomposition and different gap magnitudes may be suppressed, thus altering QPI intensity in parts of $\vec{q}$-space. Third, the normal-state anisotropy of each band strongly impacts the density of states and thus QPI intensity in $\vec{q}$-space. Here we would like to point out that one of the reasons for the successful application of the Octet / JDOS model as described below is due to the nature of the coherence factors. Particularly if for a given $\vec{k}$-point the $\Delta_i(\vec{k})=E_{B,i}(\vec{k})$, with $E_{B,i}(\vec{k})$ being the Bogoliubov quasiparticle dispersion, then the coherence factors are equal $|u(\vec{k})|^2=|v(\vec{k})|^2 = 1/2$. This leads to the intensity of scattering events connecting these points being enhanced since any other ratio of the coherence factors diminishes the relative intensity all other things being equal. Finally, the gap functions $\Delta_i(\vec{k})$ which are the objective here, influence the QPI through the dispersion and coherence factors of their Bogoliubov excitation spectrum.

In general it is possible to calculate the QPI response in a given system within the $\hat{T}$-matrix approach. This theoretical technique has been discussed in relation to quasiparticle interference in unconventional superconductors in a range of theoretical papers for both cuprates (18-20, 35, 36 and iron-based superconductors (21-24).

Here we present, first, the results of a numerical simulation using the T-matrix equations for a single band - as described for example by Wang and Lee (18). This approach is easily expandable to include multiband / multi-orbital effects that are important for Fe-based superconductors, as has been demonstrated for example by Zhang *et al.* (21). An aspect of their results of particular relevance for the experiments reported here is that there can be strong suppression of those quasiparticle interference peaks due to scattering connecting bands of different orbital character. This is consistent with the absence of strong interband scattering signals in our data.

## $\hat{T}$-matrix Simulation

In the following we briefly summarize the main equations governing quasiparticle interference in the $\hat{T}$-matrix approach following (18). Within it, the change $\delta n$ in the local density of states of a superconductor relative to the underlying unperturbed electronic liquid is given by

$$\delta n(\boldsymbol{q}, \omega) = -\frac{1}{2\pi i}[A_{11}(\boldsymbol{q}, \omega) + A_{22}(\boldsymbol{q}, -\omega) - A_{11}^*(-\boldsymbol{q}, \omega) - A_{22}^*(-\boldsymbol{q}, -\omega)]$$

(Eq. S1)

where $A_{\alpha\beta}$ are the elements of the 2 x 2 matrix $\hat{A}$ given by

$$\hat{A} = \int \frac{d^2k}{(2\pi)^2}\{\hat{G}_0(\boldsymbol{k}+\boldsymbol{q}, \omega)\hat{T}(\omega)\hat{G}_0(\boldsymbol{k}, \omega)\}$$

(Eq. S2)

with $\hat{G}(\boldsymbol{k},\omega)$ being the 2 x 2 Green's function of the Bogoliubov quasiparticle spectrum at momentum $\boldsymbol{k}$ and energy $\omega$ and $\hat{T}(\omega)$ is the scattering matrix[1]. $\hat{G}(\boldsymbol{k},\omega)$ is given by

$$\hat{G}^{-1}(\boldsymbol{k}, \omega) = (\omega + i\delta)\hat{I} - \varepsilon_k\sigma_3 - \Delta_k\sigma_1$$

(Eq. S3)

where $\sigma_i$ are the Pauli matrices, $\varepsilon_k$ the band dispersion and $\Delta_k$ the momentum-dependent gap function. The latter two are constraint by our experimental data in the following calculations[2].

We evaluated (Eq.S3) numerically on a 2000 x 2000 point grid in the positive quadrant of the tetragonal unit cell which we subsequently down-sampled to 250 x 250 points in order to minimize numerical errors due to singularities in $\hat{G}(\boldsymbol{k}, \omega)$. This was carried out for 81 energy layers between -10 meV and +10meV.

$\hat{T}(\omega)$ is a 2 x 2 matrix that can be expressed as

$$\hat{T}(\omega)^{-1} = (V_s\sigma_3 - V_m\hat{I})^{-1} + \int \frac{d^2k}{(2\pi)^2}\hat{G}_0(\boldsymbol{k}, \omega)$$

(Eq. S4)

with $V_s$ being a non-magnetic and $V_m$ being a magnetic scattering potential. For the simulations reported below we have chosen $V_s$ ($V_m$) - the only parameter in this framework not directly constraint by our experiments - to be of the order of 100 meV – a value consistent with previous work (20,23). The results of the numerical evaluation of (Eq.S3) and (Eq.S4) were then used to evaluate (Eq.S1) and (Eq.S2).

---

[1] Here we already introduced the generally made approximation that $\hat{T}(\omega)$ is momentum independent – a characteristic of a point like *s*-wave scatterer.
[2] Throughout the calculation we have set the energy broadening term δ to 0.0003 - a small but finite value consistent with general practice in the literature. See for example (23).

In Figure S3 we summarize the calculations by showing the resulting $g(\vec{q}, E)$ for $\hat{I}$ (Fig. S3A) and $\sigma_3$- scattering (Fig. S3C). From the identical simulations, we show in Fig. S3B,D the scattering intensity in the $\overrightarrow{|q|} - E$ plane along the same high symmetry $\Gamma\widetilde{M}$ direction as in the main text. The basic 'clamshell' structure seen so clearly here (and in all types of simulations) is generated by the gap-energy dependence of two octet scattering vectors, $\vec{q}_3$ and $\vec{q}_7$ (see Fig. 1E,F for details), that always lie in the high symmetry $\Gamma\widetilde{M}$ plane for this band. This particle-hole symmetric 'clamshell' of scattering intensity within the $\overrightarrow{|q|} - E$ plane is the hallmark of the anisotropic superconducting energy gap on that band, and this realization is central to the new multi-band QPI technique we wish to report. It should be pointed out that in order to extract the information contained in the 'clamshell' structure the data has to be analysed differently to the traditional single $\vec{q}$-layer analysis by evaluating cuts in the $\overrightarrow{|q|} - E$ plane. Though this is simply another way of slicing our 3 dimensional data sets it is experimentally far more challenging since it requires not just a select number of $|\vec{q}|$ maps but a complete set as a function of energy.

The experimental $g(\vec{q}, E)$ data most probably represent a mixture of the two types of scattering ($\sigma_3$ and $\hat{I}$) due to a range of scatterers contributing to the overall signal. The most significant difference between the two cases is the occurrence of additional scattering events close to the gap edges for $\sigma_3$-scattering; a similar effect of an intense, broad scattering centered at Γ and at energies close to the gap edge has been observed by Zhang *et al.* (23).

*JDOS Simulation*

In Figure S3E and S3F we show equivalent simulations but using the Joint-Density-of-States approximation:

$$\mathrm{JDOS}(\boldsymbol{q}, \omega) = \int \frac{d^2 k}{(2\pi)^2} \{\mathrm{Imag}(\hat{G}_{11}(\boldsymbol{k} + \boldsymbol{q}, \omega) \times \mathrm{Imag}(\hat{G}_{11}(\boldsymbol{k}, \omega))\}$$

(Eq. S5)

The key observation is that, in terms of the geometrical characteristics of $g(\vec{q}, E)$, the JDOS simulation is virtually indistinguishable from that of the $\hat{T}$-matrix approach. As can be seen once again, the quasiparticle dispersion as well as the 'clamshell feature' are prominent as points of maximum intensity in the $\vec{|q|} - E$ plane. The JDOS approximation is based on the observation that the integral in (Eq. S2) is dominated by terms where the Green's functions at **k** and **k+q** are both simultaneously large, which happens in general only in the immediate vicinity of the poles, i.e. singularities of the imaginary part. In other words, from the most simplest point of view $\delta n(\boldsymbol{q}, \omega)$ will be large if **q** connects regions of large density of states for the unperturbed Green's function.

In the context of data presented in the paper, the most important observation in Fig. S3A-F is that, for both $\hat{T}$-matrix and JDOS cases, the $g(\vec{q}, E)$ features upon which we base our analysis (quasiparticle dispersion plus the intense arc of scattering for energies $\Delta^{\min} \leq E \leq \Delta^{\max}$) are prominent and do not depend significantly on the details of the scatterer.

*FeAs Octet Model*

Finally, we comment on the relation of the above two approximation schemes to the well-known octet model. For this purpose we have overlayed on the simulated $g(\vec{q}, E)$ in Figs S3A,C , the positions of the 8 singular vectors as predicted by the octet model (main text Fig. 1). As can be seen by comparison, these positions correspond very well indeed to special high scattering intensity vectors in both the JDOS and $\hat{T}$-matrix approximation for $g(\vec{q}, E)$.

In particular, it becomes apparent that the arc of maximal scattering intensity predicted by both T-matrix and JDOS simulations (which is key to the empirical procedure for determination of $\Delta_i(\vec{k})$ in main text) is related to the dispersion of the two specific octet $\vec{q}$-vectors $\vec{q}_3$ and $\vec{q}_7$; we emphasize this using larger red dots in Figures S3A,C,G. These vectors are required by symmetry to lie on the diagonal of the Brillouin zone and in a high symmetry plane of the anisotropic gap structure. This leads to the geometric structure of scattering intensity maxima in the $\vec{|q|} - E$ plane shown in Fig. S3G,H. We discuss the relation of the octet model to the extraction of the gap anisotropy in more detail in the following section of the SOM.

*Summary*

In summary, we have discussed the expected QPI-response of a hole-like band (as observed in LiFeAs) on which an anisotropic superconducting gap has opened. We have shown that, on each level of approximation (octet model, JDOS and $\hat{T}$-matrix), quasiparticle interference leads to experimentally indistinguishable geometrical characteristics in $g(\vec{q}, E)$. All three approaches to understanding $g(\vec{q}, E)$ also show directly that the *geometric* information of the *position* of the maxima in $\overrightarrow{|q|} - E$ space encode sufficiently constrained information on both the band structure $\varepsilon(\vec{k})$ and the anisotropic gap structure $\Delta(\vec{k})$. Therefore the geometric information on which our analysis is based has a sound footing, not only in the intuitive octet model and JDOS approximation, but also the $\hat{T}$-matrix approximation.

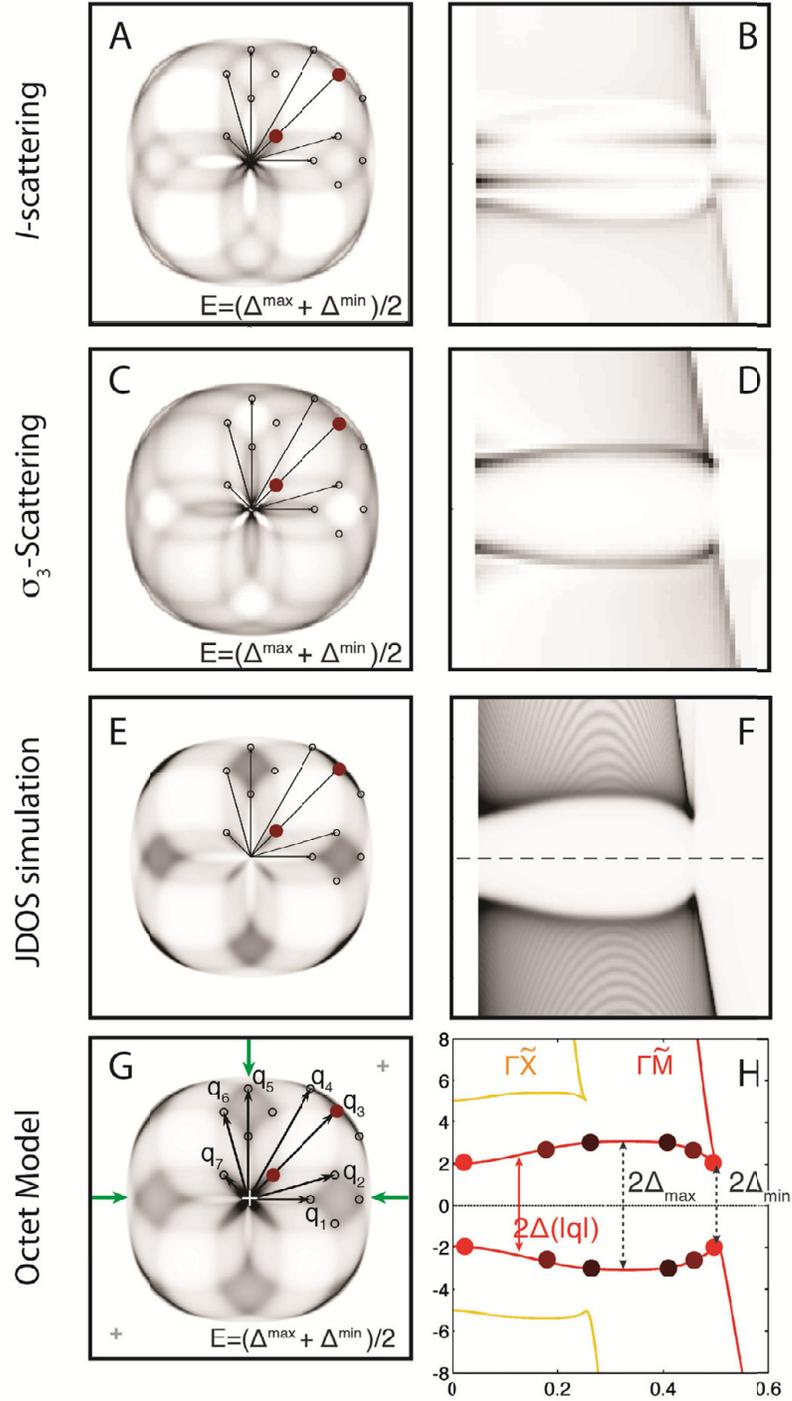

**Fig. S3** Results of T-matrix simulation (panels A-D), the JDOS simulation (panel E,F), and the octet model (panel G,H). For details see text.

# III. Quantitative Determination of the Superconducting Gaps $\Delta_i(\vec{k})$ from the Scattering Intensity Maxima in the $|\vec{q}|$-E plane (Materials and Methods B)

As discussed in the previous section of the SOM, the curved arc of maximal scattering intensity along the high symmetry direction in $\vec{q}$-space contains key information on $\Delta_i(\theta)$, the dependence on angle θ in $k$-space of the energy gap for a given band *i*. Here we make explicit the procedure to invert the measured scattering intensity along the specific high-symmetry $|\vec{q}| - E$ plane to obtain the momentum resolved superconducting energy gap $\Delta_i(\vec{k})$ for each band, making clear that this approach is independent of the $\hat{T}$-matrix or JDOS simulations.

Our starting point is the Bogoliubov quasiparticle excitation spectrum. In Fig. S4A we show contours-of-constant-energy *E* for a Bogoliubov quasiparticle spectrum of an anisotropically gapped band at energies $\Delta_i^{max} > E > \Delta_i^{min}$. Central to the discussion are the points demarking the edges of the banana-shaped constant-energy-contours. Due to the high density of states and the coherence factors associated with them the QPI signature at a given energy will be centered on signals connecting these points ('octet-model', see previous section of SOM).

The ($k_x$, $k_y$) coordinates of these points of most intense scattering lie on the same contour as the Fermi surface of the underlying normal state band structure, as indicated in Fig. S4B. The reason for this is that the superconducting gap opens particle-hole symmetric relative to the Fermi surface. By extracting the Fermi surface from our band structure fits, we are therefore able to determine the contour of high-intensity scattering points in the $k_x/k_y$ plane.

Next we consider all scattering vectors along the high symmetry direction $\Gamma\widetilde{M}$ connecting these points (Fig. S4C). As can be seen for each set of eight equivalent points on the Fermi surface there exist exactly two QPI vectors along the $\Gamma\widetilde{M}$ direction connecting them. Most importantly, there is a direct correspondence between the length $|\vec{q}|$ of such a vector and the magnitude of the gap value of the octet model it belongs to, i.e. a certain vector length corresponds to only one $\Delta(|\vec{q}|)$. Plotting these vectors in a $|\vec{q}|$-*E* plot leads to the curved shape shown in Fig. S4D and Fig. 1H of the main text.

In Fig. S4E we show how the length of each of these vectors can be directly translated into an angle. Firstly,

$$\theta = \operatorname{atan}\left(\frac{k_{y1}}{k_{x1}}\right),$$

(Eq. S7)

where ($k_{x1}$, $k_{y1}$) are the coordinates of point (1) on the Fermi surface. Secondly as shown in the same figure, one can calculate the magnitude $|\vec{q}|$ of the QPI scattering vector connecting points (1) and (2) as

$$\begin{aligned} |\vec{q}| &= \sqrt{(k_{x1} - k_{x2})^2 + (k_{y1} - k_{y2})^2} \\ &= \sqrt{2}|(k_{x1} - k_{x2})| \\ &= \sqrt{2}|(k_{x1} + k_{y1})| \end{aligned}$$

(Eq. S8)

Here we used that on geometric grounds $k_{x1}$=-$k_{y2}$ and $k_{y1}$=-$k_{x2}$. As mentioned earlier, the points ($k_x$, $k_y$) have to lie on the Fermi surface that we can describe numerically based on a band structure fit of the dispersion extracted from our data. For each coordinate ($k_x$, $k_y$) we can therefore calculate a triplet ($|\vec{q}|$, θ, Δ). Due to the $C_4$ symmetry of the problem and the high symmetry direction of the analysis each $|\vec{q}|$ vector corresponds to only one gap value Δ. This allows to establish an exact correspondence between ($k_x$, $k_y$) and (θ, $|\vec{q}|$) where all θ leading to the same $|\vec{q}|$ have equivalent gap values in a $C_4$ symmetric system.

For completeness, we show that it is possible to get a closed expression in case of a circular Fermi surface described by a single Fermi wave vector $k_F$ (Fig. S3F). In this case simple geometric consideration results in

$$\begin{aligned} |\vec{q}|/2 &= k_F \times \sin(\alpha) \\ &= k_F \times \sin(135 - \theta) \\ &= k_F \times \cos(\theta - 45). \end{aligned}$$

(Eq. S9)

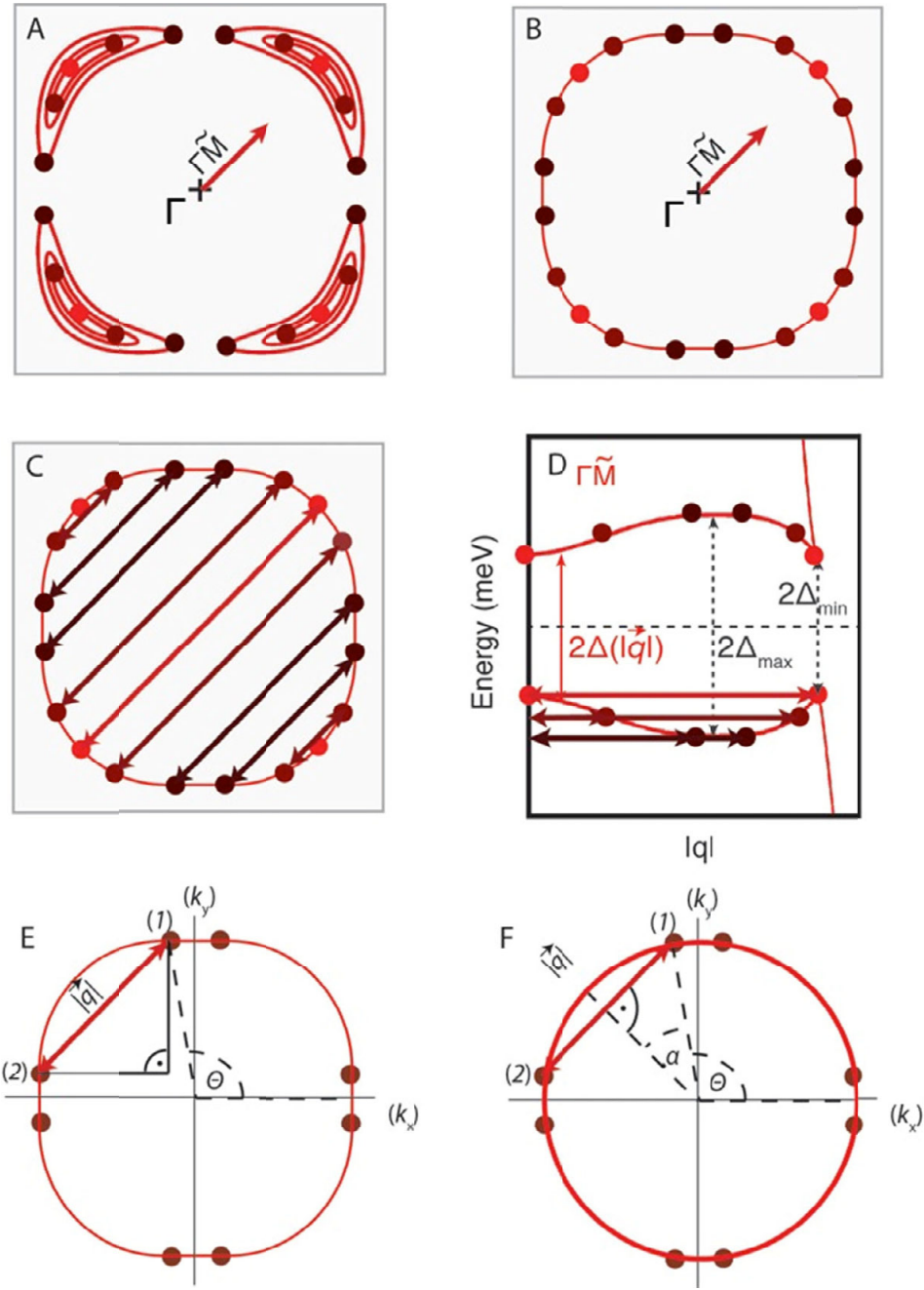

**Figure S4 | Angle-to-momentum conversion.** A, Schematic contours-of-constant-energy for one of the two anisotropic gaps shown in Fig. 1C. Contours enclose diminishing areas surrounding the gap minimum in each case. The dots indicate points of strong density of states for each contour. B, The points are the same as in Fig. S4A. Due to particle-hole symmetry of the superconducting state they lie on the contour defined by the Fermi surface of the normal state Fermi liquid quasiparticle band structure. C, In addition to B, the arrows are indicating the dominant quasiparticle scattering along the high symmetry direction $\Gamma\tilde{M}$ at the energy corresponding to the closed contours in A. D, The resulting curved shape by plotting the vectors from C in a $|\vec{q}|$-$E$ plot. E, Schematic diagram explaining the correspondence between the angle θ and the length $|\vec{q}|$ of the scattering vector dominant in the direction $\Gamma\tilde{M}$ at the energy equal to the gap Δ(θ). F, Angle-to-momentum conversion for a circular Fermi surface.

## IV. Topographic Image

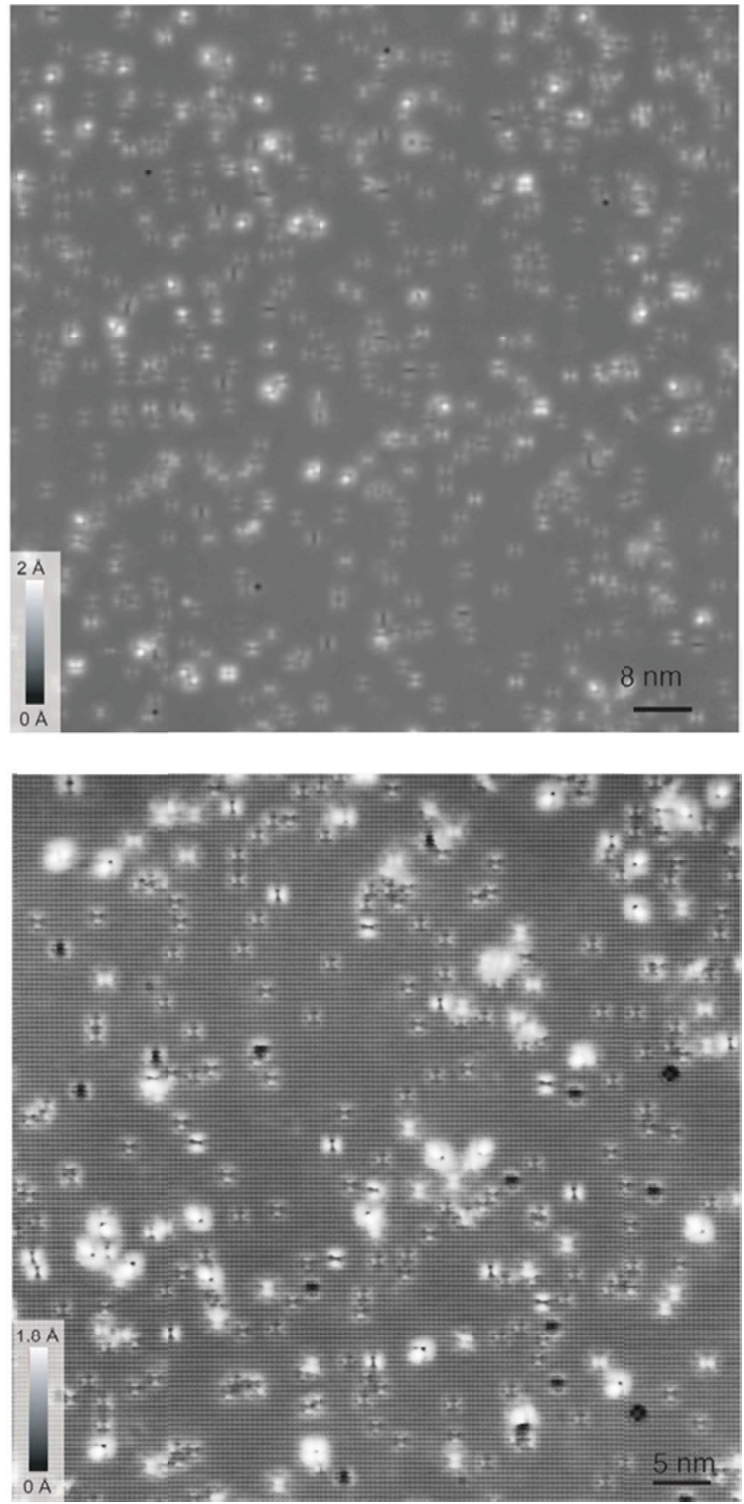

**Figure S5 | Topographic images of the LiFeAs surface. A**, Typical topography taken simultaneously with a $g(|\vec{r}|, E)$ measurement, using $V_{setup}$=-50meV, $I_{setup}$=10pA. **B**, Low junction resistance topography ($V_{setup}$=-30meV, $I_{setup}$=30pA) that shows clear atomic corrugation.

## V. Characteristics of In-Gap States

The spectra in clean regions of the LiFeAs surfaces show a flat bottom with no states between ±2meV, except residual spectral weight consistent with thermal broadening (Fig. 3A, inset). However, we find a distinct set of impurities on the LiFeAs surface that exhibit an in-gap state (Fig. S6A). Close to such impurities, spectral weight is shifted to ±1.5meV. These impurities lie on or above the Fe site, and their topographic signature is shown in Fig. S6B. The local density of states at the energy of the in-gap state shows a distinct pattern with lobes extending the Fe-Fe direction, and a clear wave pattern around it with a wavelength similar to the Fermi vector of the $h_2$ band (Fig. S6C). The pattern is $C_2$ symmetric, probably because of the local $C_2$ symmetry of the Fe site. The Fourier transform $g(|\vec{q}|, E)$, shows a clear signal from the in-gap state. When averaged over all $C_2$-symmetric impurities, an effectively $C_4$ pattern emerges as shown in Fig S6D. Finally, in Fig. S6E we show a set of spectra taken through the in-gap-state along the horizontal of C. The effects of these in-gap bound states are excluded from the QPI analysis.

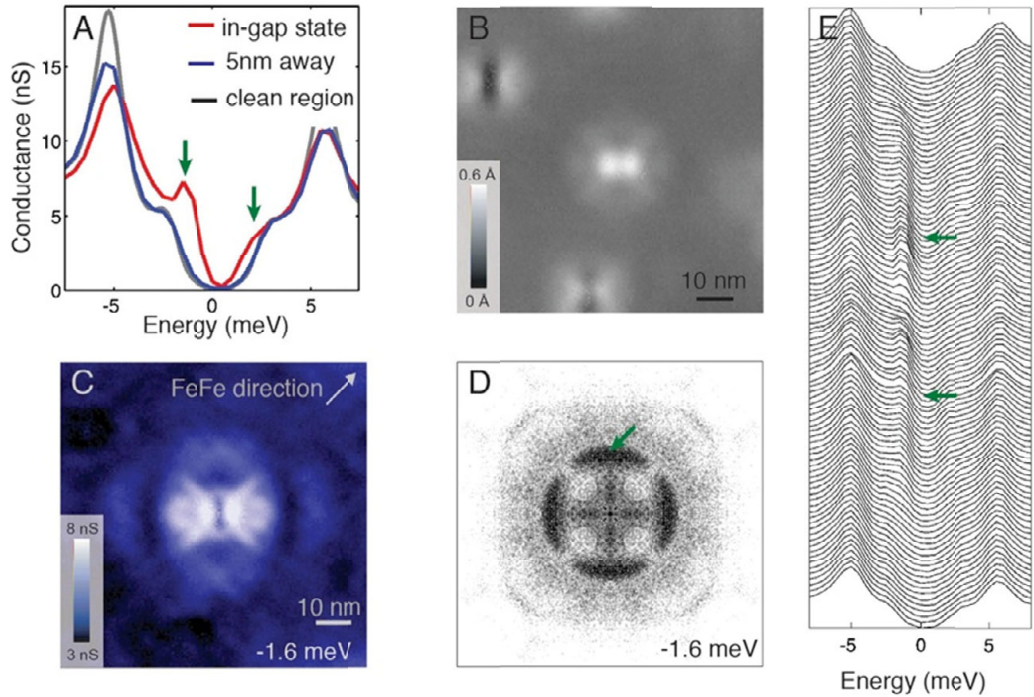

**Figure S6 | In-gap state.** A, Spectra in clean regions and at the impurity locations that exhibit in-gap states. B, Topograph, with such an impurity at the center. C, The local density of states ~$g(\vec{q}, E=1.6\text{meV})$ at the same location, showing the real-space shape of the in-gap state as well as Friedel-like oscillations around it. D, Signature in $\vec{q}$-space (arrow). E, A set of spectra taken through the in-gap-state along the horizontal of C. The arrows indicate the in-gap state.

# VI. Sequence of $g(\vec{r}, E)$ Conductance Images and their $g(\vec{q}, E)$

In Fig. S7, the right column shows a sequence of raw $g(\vec{r}, E)$ images, the center column shows their respective PSD Fourier transforms $g(\vec{q}, E)$. The left column are the same images after simple background subtraction as described below. The green arrows are as in the main text and indicate the suppression of scattering for h$_2$, h$_3$ due to the opening of the anisotropic superconducting gap in those directions. The blue arrow indicate $g(\vec{q}, E)$ intensity that stems from the in-gap state located at certain impurity locations and visible only in their vicinity, as described in Section V.

Unprocessed $g(\vec{q}, E)$ images show a peak around $\vec{q}$ =(0,0). This peak stems from long range spatial variations of the surface and from randomly scattered defects; it is not related to quasiparticle scattering. To obtain better visibility, we therefore suppress the intensity of very small $\vec{q}$-vectors around the center: QPI$_{new}$=QPI$_{raw}$[1–*Gaussian*($\vec{q}$ (0,0),σ)]. Last, we symmetrized the data to obtain images shown in the left column. The center column shows the raw data.

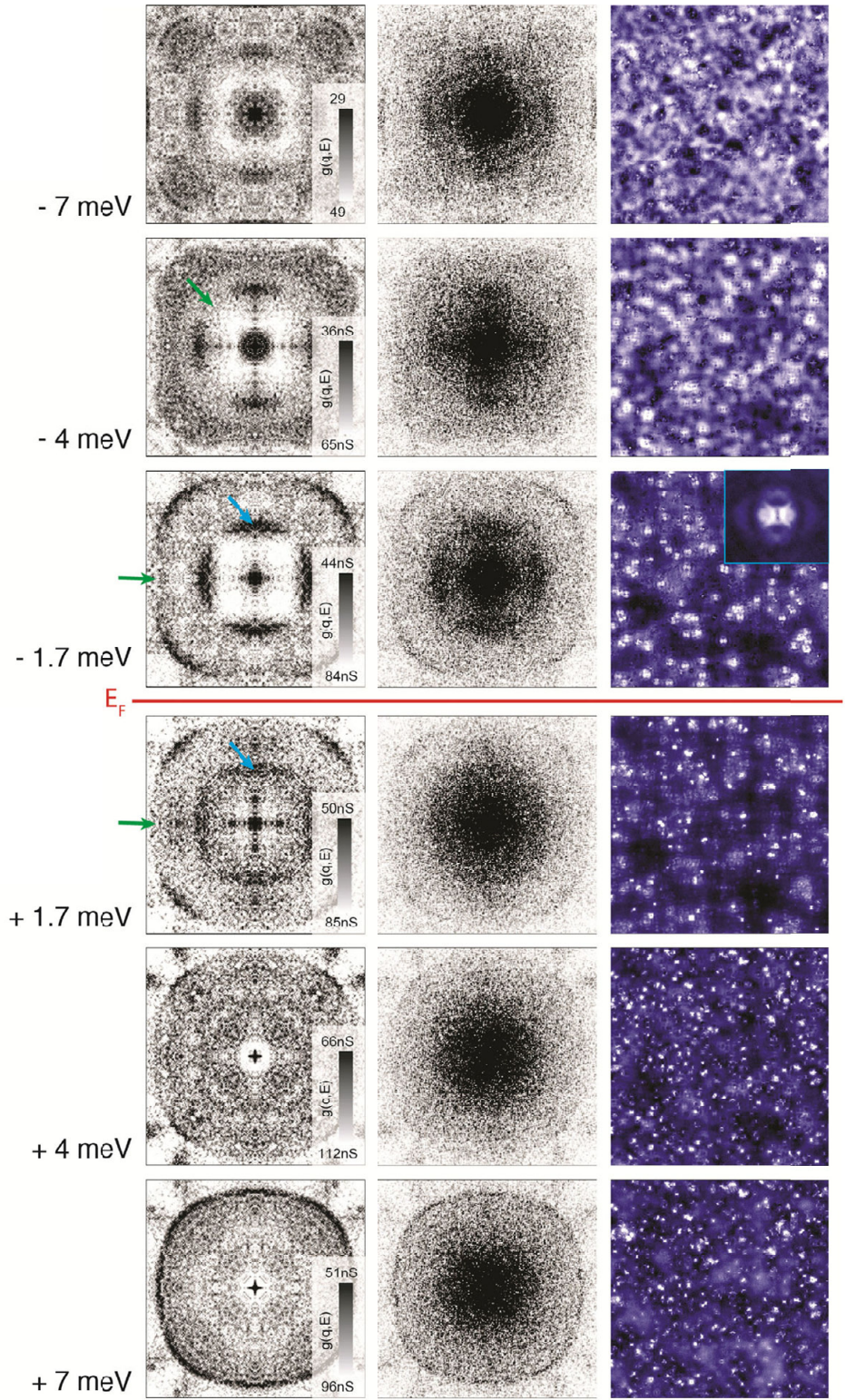

**Figure S7 |** Sequence of $g(\vec{r}, E)$ (right column) and their respective Fourier transforms $|g(\vec{q}, E)|$ (left column). The center column shows the raw Fourier transforms, the left column is processed as described in the text. The inset at -1.7meV shows an enlarged FOV of an exemplary impurity bound state that exists at that energy and is described in more detail in section V.

## VII. Quantitative Comparison with Quantum Oscillation

Here we briefly demonstrate the consistency of the dispersions identified by us with the available bulk Fermi surface structure from quantum oscillations as published by Putzke *et al.* (31).

The quantity that can be compared most directly is the size of the Fermi surface. In case of quantum oscillations it is given directly by the frequency of the observed oscillations.

In Fig. S10 we reproduce part of Fig. 2c in (31) with kind permission of the authors. Here the black squares show the measured frequency $F$ as a function of angle $\Theta$. The colored lines are based on a spin-orbit coupling DFT calculation that was adjusted in the right panel in order to best describe the experimental quantum oscillation data (31).

One can compare the quantum oscillation data directly to the size of the Fermi surface as obtained from our QPI measurements. Based on the tight binding fits we evaluated the size of $h_3$ and $h_2$ to be 16.1% and 4.9% of the Brillouin zone corresponding to a dHvA frequency of 4.66kT and 1.41kT respectively ($h_1$ does not conclusively cross the Fermi surface in our data and the corresponding band was neither observed in dHvA). We show these frequencies as blue dots in Fig. S10.

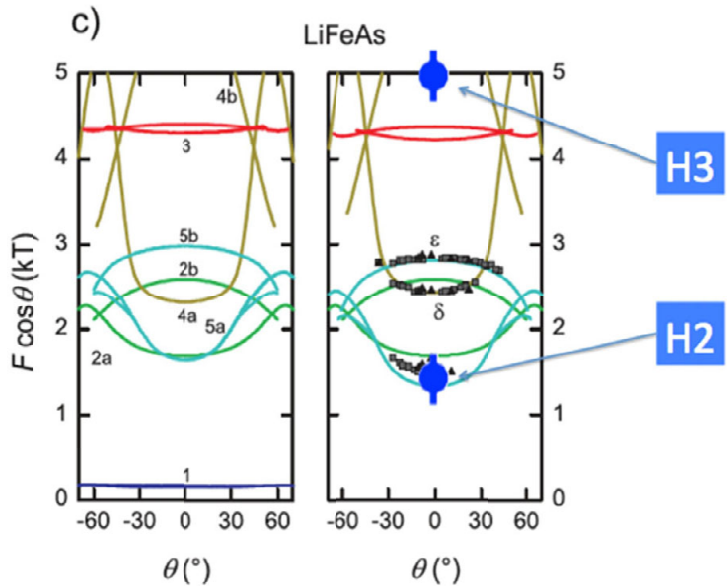

**Figure S8 |** Comparison of Fermi surface size extracted from our data (blue dots) with the data and DFT calculation by Putzke et al. Left Panel: DFT calculation. Right Panel: DFT calculated band energies slightly adjusted to best fit the quantum oscillation data. (Figure reproduced with kind permission by authors of ref. (31)).

As can be seen the Fermi surface size for $h_3$ is in good agreement with the calculated band 3 (red) which is the γ band, whereas $h_2$ is consistent with the observed data, which can be assigned to be either band 2 (a hole band) or band 5 (an electron band) in the DFT calculation by Putzke *et al.* While the authors of ref. (38) believe their observed frequency to be dominated by an electron pocket they are not excluding the possibility of the signal to be a mixture originating from both the orbits 2a and 5a. (Quantum oscillations cannot directly determine the electron- or hole-like character of the measured frequencies). The remaining two orbits measured by Putzke *et al.* are reported to be electron orbits of which we do not observe a strong signature. Furthermore a reason for the strong suppression of the hole like orbit 2b in our data is most probably its different (finite) $k_z$ value.

# VIII. Extraction of the Positions of Maximum Scattering intensities in the $|\vec{q}| - E$ plane

### VIII.1. $h_3$ Band

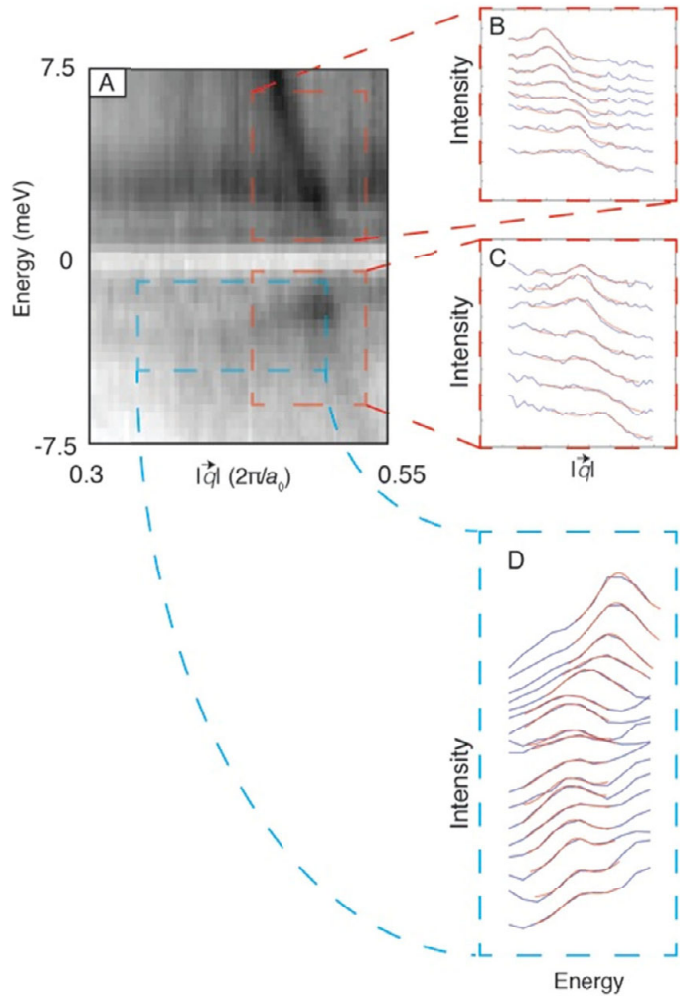

**Figure S9** | Extracting the peak positions of the $h_3$ band and the curved signal that stems from the anisotropic gap that are used in Fig. 3D. A, Intensity plot of a $|\vec{q}|$-E cut along $\Gamma\widetilde{M}$, the minimum gap direction of h3. B, C, Line traces of scattering intensity as a function of momentum for fixed energies from the data shown in the red rectangle in A. The red lines are fits to the data using a Gaussian function on a second-degree polynomial background with the center of the Gaussian used to extract the $|\vec{q}|$ value of the dispersing signal. For Fig. 3D we averaged such data from 4 different measurements. D, Line traces of the scattering intensity as a function of energy for fixed momenta. The position of the curved line is extracted the same way as described for B, C.

*VIII.2. h₂ Band*

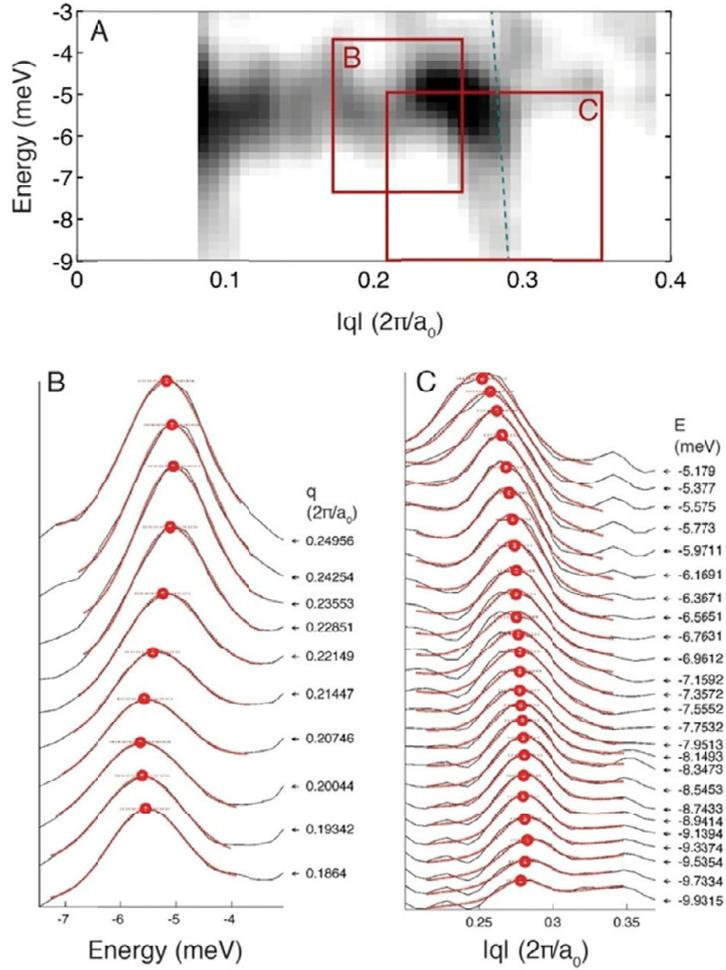

**Figure S10 |** Same as Fig. S8, but for the h₂ band. A, Intensity plot of a $|\vec{q}|$,E cut along $\Gamma\tilde{X}$, the minimum gap direction of h₂ (a background has been subtracted as in Section VI). The blue line is a guide to the eye along the normal state h₂. B, Constant energy traces from the data shown in the red rectangle in A. The red lines are fits to the data using a Gaussian function on a second-degree polynomial background; the red dots indicate the maxima. For Figure 3D, we averaged such data from four different measurements. C, Same as in B but with constant $\vec{q}$-traces.

22